\documentclass[aps,prd,preprint,nofootinbib]{revtex4}
\usepackage{amsfonts}
\usepackage{mathrsfs}
\usepackage{graphicx}
\usepackage{caption}
\usepackage{subcaption}
\usepackage{amsmath}
\usepackage{amssymb}
\usepackage{epsfig}
\usepackage{graphicx}
\usepackage{slashed}
\usepackage{color}
\usepackage{multirow}
\usepackage{ulem}
\usepackage{adjustbox}

\usepackage{stmaryrd}
\allowdisplaybreaks[1]

\graphicspath{{fig/}{dia/}} \DeclareGraphicsExtensions{.eps}
\begin{document}

\title{New Signature of  low mass $Z^\prime$ in $J/\psi$ decays}

\author {Chao-Qiang Geng$^1$, Chia-Wei Liu$^2$ and Jiabao Zhang$^{1,3}$\footnote{zhangjiabao21@mails.ucas.ac.cn}}
\affiliation{
$^1$School of Fundamental Physics and Mathematical Sciences, Hangzhou Institute for Advanced Study, UCAS, Hangzhou 310024, China\\
$^2$Tsung-Dao Lee Institute \& School of Physics and Astronomy, Shanghai Jiao Tong University, Shanghai 200240, China\\
$^3$Institute of Theoretical Physics,~UCAS, Beijing 100190, China\\
University of Chinese Academy of Sciences, 100190 Beijing, China
}
\date{\today}

\begin{abstract}
We explore a new approach to search for a low-mass $Z^{\prime}$ particle through $J/\psi$ decays by identifying its existence through parity-violating phenomena in the isospin-violating final states of $\Lambda\overline{\Sigma}^{0}$ and the corresponding charge conjugated states of $\overline{\Lambda}\Sigma^{0}$. 
Our investigation centers on a generation-independent and leptophobic $Z^{\prime}$ with its mass below 10 GeV.
Given the present experimental conditions at the Beijing Spectrometer III~(BESIII) and the anticipated opportunities at the Super Tau Charm Factory~(STCF), we conduct Monte-Carlo simulations to predict possible events at both facilities.
%{\color{red}Our simulations indicate that BESIII experiments hold the potential to detect $Z^{\prime}$ signals in $J/\psi\to\Lambda\overline{\Sigma}^{0}$ if the polarization asymmetry paramter $\alpha_{\text{NP}}$ attains a minimum threshold of 0.02, and the systematic uncertainty are further reduced.}
Notably, we foresee a substantial enhancement in the precision of the lower limit estimation of $\alpha_{\text{NP}}$ as well as a reduction in statistical uncertainty with upcoming STCF experiments.
Furthermore, it is essential to highlight that a null result in the measurement of $\alpha_{\text{NP}}$ would impose stringent constraints,  requiring the $Z^{\prime}-q-q$ couplings to be on the order of $10^{-2}$. 
\end{abstract}

\maketitle

% \section{Introduction}

As an extra neutral U(1) gauge boson, $Z^{\prime}$ manifests itself in many extensions of the standard model~(SM), such as the Grand Unified Theories~(GUTs)~\cite{Barr:1980fg,Robinett:1981yz,Langacker:1984dc,London:1986dk,DelAguila:1995fa,Hewett:1988xc}, heterotic string theory~\cite{Komachenko:1989qn}, left-right symmetric models~\cite{Mohapatra:1974hk,Barger:1978rj,Ma:1986we,Babu:1987kp}, and gauged B-L models~\cite{Davidson:1978pm,Li:1982gc,Basso:2008iv}.
Searching for such a gauge boson helps us to gain more insights about the fundamental theory beyond the SM.
Experimentally, direct searches for the $Z^\prime$ boson are conducted in various types of high energy colliders, including $e^{+}e^{-}$ colliders like LEP, and hadron colliders such as Tevatron and LHC.
Various mass ranges of $Z^{\prime}$ are scanned, and the couplings of $Z^{\prime}$ with both leptons and quarks are constrained.
In the case of leptonic collider searches, the agreement between LEP-II measurements and the SM predictions regarding the cross-section of $e^{+}e^{-}\to f\bar{f}$ implies that either $M_{Z^\prime} > 209\,$GeV, or that the $Z^\prime$ couplings with leptons are smaller than $10^{-2}$~\cite{Appelquist:2002mw, Carena:2004xs, ALEPH:2006bhb}.
Stronger constraints have also been found through the dark photon searches in various experiments~\cite{ATLAS:2019erb,CMS:2015nay,LHCb:2019vmc,BaBar:2014zli}. The limit on the couplings between $Z^{\prime}$ and leptons is almost around $10^{-4}$.
Besides, some indirect searches through neutrino-electron scatterings are also proposed and severe constraints are given various neutrino experiments~\cite{Chakraborty:2021apc,Asai:2022zxw}.

These searches have led to the consideration of the leptophobic $Z^\prime$ boson, which interacts exclusively with quarks and is extensively searched on hadronic colliders.
Through extensive scanning of the dijet mass spectrum, upper limits on the $Z^{\prime}$ couplings have been established by the CMS collaboration in the mass range from several TeV down to 10 GeV~\cite{CMS:2016ltu, CMS:2018ucw, ParticleDataGroup:2022pth}.
For $Z^\prime$ bosons with masses below 10 GeV, comprehensive explorations on the hadron colliders are limited due to significant background interferences.
While some progress has been made through nonstandard quarkonium decays~\cite{Dobrescu:2014fca}, there remains a pressing need for additional strategies to comprehensively investigate this specific low mass range.

In addressing this critical research gap, we propose to conduct the search of the $Z^{\prime}$ boson on lepton colliders, such as Beijing Spectrometer III~(BESIII) and the forthcoming Super Tau Charm Factory~(STCF), which have a very clean background as well as large volumn of data sample.
The BESIII collaboration achieved a significant milestone, accumulating a staggering $10^{10}~J/\psi$ events by 2019, with considerable amount of events producing polarized baryon-antibaryon pairs~\cite{Mangoni:2020nsc}.
Utilizing the entanglement of final states has enabled the extraction of observables at an unprecedented level of accuracy, offering an excellent platform for probing new physics~(NP) phenomena~\cite{He:2022jjc, Fu:2023ose}.
Moreover, the future STCF is designed to take $1ab^{-1}$ data, corresponding to $3.4\times10^{12}~J/\psi$ events per year~\cite{Zeng:2023wqw}, promising even higher precision in relevant processes.
One of the $Z^{\prime}$ models has been proposed to relieve the tensions in the $J/\psi\to\pi^{+}\pi^{-}$ and $\psi(2S)\to\pi^{+}\pi^{-}$ branching fractions with fitted pion form factors~\cite{Bause:2022jes}.
In this work, we focus on parity violation in $J/\psi\to\Lambda\overline{\Sigma}^{0}$ and its charge conjugate.
Dominated by a single virtual photon exchange, the nonperturbative effects stemming from gluon exchanges in such decays are comparatively suppressed, allowing for a factorizable amplitude at the first order in theoretical calculations~\cite{Geng:2023yqo, BESIII:2012xdg}.
Furthermore, the SM prediction for parity violation in $J/\psi \to \Lambda \overline{\Sigma}^0+c.c.$ is vanishingly small, leading to a clean background for the detection of $Z^{\prime}$.
The BESIII collaboration has very recently analyzed CP violation in $J/\psi \to \Lambda\overline{\Sigma}^{0}$~\cite{BESIII:2023cvk}. They have measured the ratio between electric and magnetic form factors with high precision, demonstrating their capability to accurately reconstruct decay distributions. However, their work assumed spatial inverse parity symmetry, a constraint we have relaxed in our research, representing the main contribution of our study.

% \section{One-fold angular distributions}

The parity violating effect is characterized by the polarization asymmetry parameter $\alpha_{\text{NP}}$ for the decay of $J/\psi \to \Lambda\overline{\Sigma}^0$.
Experimentally, $\alpha_{\text{NP}}$ is available from the angular distribution as follows:
\begin{equation}\label{eq1}
\frac{1}{\Gamma}\frac{\partial \Gamma }{\partial \cos \theta_p}=1+\alpha_{\text{NP}}\alpha_\Lambda\cos\theta_{p}\,=1+\alpha\cos\theta_{p}\,,
\end{equation}
where $\alpha_\Lambda = 0.748(7)$ is the asymmetry parameter in $\Lambda \to p \pi^-$~\cite{ParticleDataGroup:2022pth}, and $\theta_p$ is the angle between $\vec{p}_\Lambda$ and $\vec{p}_p$ defined at the rest frames of $\Lambda$, respectively.
Theoretically, $\alpha_{\text{NP}}$ is defined as
\begin{equation}\label{eq2}
\alpha_{\text{NP}} = \frac{|h_{++}|^2 + |h_{+-}|^2 - |h_{-+}|^2 - |h_{--}|^2}{|h_{++}|^2 + |h_{+-}|^2 + |h_{-+}|^2 + |h_{--}|^2}\,,
\end{equation}
where $h_{\lambda \overline{\lambda}}$ are the helicity amplitudes of $J/\psi \to \Lambda \overline{\Sigma}^0$
with $\lambda $ and $\overline{\lambda}$ the helicities of $\Lambda$ and $\overline{\Sigma}^0$, respectively.
When parity symmetry holds, we have $|h_{\lambda \overline{\lambda} }|^2 = |h_{-\lambda -\overline{\lambda} }|^2$ and consequently $\alpha_{\text{NP}}=0$.
Therefore, a nonzero value of  $\alpha_{\text{NP}}$ indicates the violation of parity symmetry.

\begin{figure}
\includegraphics[width=.5\linewidth]{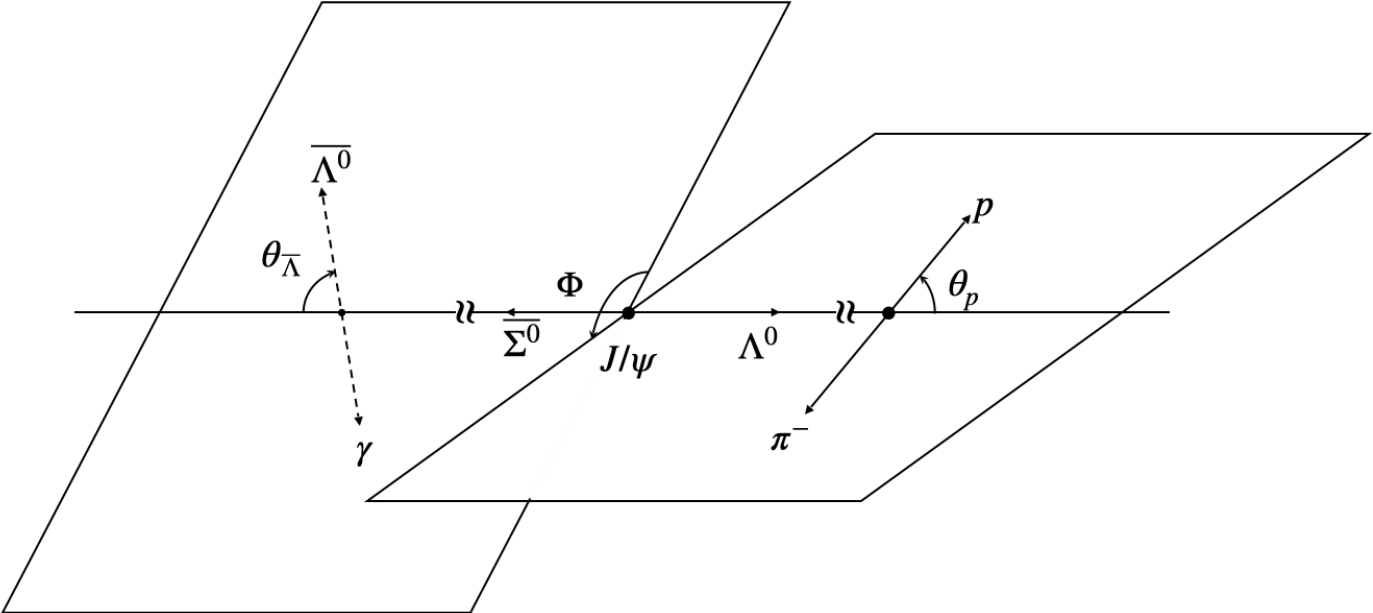}  
\caption{Decay distributions of $J/\psi\to\Lambda(\to p\pi^{-})\overline{\Sigma}^{0}(\to\overline{\Lambda}\gamma)$.}
\label{angular}
\end{figure}

Additionally, the angular distribution for the charge-conjugate process, namely, $J/\psi \to \Sigma^0 \overline{\Lambda}(\to \overline p \pi ^ + )$, is given by simply substituting $(\overline \alpha_{\text{NP}},\overline \alpha_\Lambda,\overline \alpha)$ for $(\alpha_{\text{NP}},\alpha_\Lambda, \alpha)$ in Eq.~\eqref{eq1}.
It is important to note that $\overline\alpha_\Lambda$ denotes the asymmetry parameter for $\overline{\Lambda} \to \overline p \pi^+$, with a measured value of $-0.757(4)$ according to the Particle Data Group~(PDG)~\cite{ParticleDataGroup:2022pth}.
Under the assumption that CP symmetry is conserved in the decay of $\Lambda\to p\pi^-$, we construct the CP-even and CP-odd observables as $\alpha_{\pm}=(\alpha\pm\overline\alpha)/2$. It is worth highlighting that within the SM, both $\alpha_+$ and $\alpha_-$ remain below the threshold of $10^{-3}$.
Furthermore, by considering two fold cascade decays, such as the case depicted in Fig.~\ref{angular}, more observables can be extracted.

We adopt the general effective Lagrangian describing the $Z^{\prime}$ boson, as prescribed by the PDG~\cite{ParticleDataGroup:2022pth}.
In the context of the $J/\psi\to\Lambda\overline{\Sigma}^{0}$ decay process, our analysis focuses exclusively on the isovector-axial vector current, denoted as $(\bar{u}\gamma_\mu\gamma_5 u-\bar{d}\gamma_\mu\gamma_5 d)$, and the vector current of $\bar{c}\gamma_{\mu}c$.
Consequently, the effective Lagrangian tailored for our investigation is as follows:
\begin{equation}\label{eq3}
 Z^\prime_\mu\left[g_A \left(\overline{u} \gamma^\mu \gamma_5 u-\overline{d} \gamma^\mu \gamma_5 d\right) +g_V \overline{c} \gamma^\mu c\right]+ {\cal C}\,.
\end{equation}
where $g_A = (g_u^R - g_d^R)/4$ and $g_V = (g_u^R + g^L)/2$ represent the pertinent coupling constants.
The vector current $\bar{c}\gamma_{\mu}c$ is dictated by the annihilation of $J/\psi$, and the axial vector currents of $u,d$ quarks are considered to introduce parity violation.
Due to the requirement of Hermiticity, $g_A$ and $g_V$ must be real, thereby ensuring CP conservation and  $\alpha_- = 0$. 
Other terms in the Lagrangian are collectively designated as ${\cal C}$, and do not affect the detection of $Z^{\prime}$.
In the presence of such a $Z^{\prime}$ boson, parity violation arises from the interference between amplitudes associated with $J/\psi \to Z^{\prime *}/\gamma^{*} \to \Lambda \overline{\Sigma}^0$.
These amplitudes are labeled as ${\cal A}_{Z^\prime/\gamma}$ and, at the first order, they are given as:
\begin{equation}\label{eq4}
{\cal A}_{Z^\prime}=2 g _A g_V f_{ \psi}M_{\psi}S_{Z^{\prime}}\epsilon_\mu\langle \Lambda \overline{\Sigma}^0|\overline{u}\gamma^\mu \gamma_5u|0\rangle\,,~
{\cal A}_{\gamma }=\frac{8}{3}\pi\alpha_{em} \frac{f_{\psi}}{M_{ \psi}} \epsilon_\mu\langle \Lambda \overline{\Sigma} ^ 0 |\overline{u} \gamma^\mu u|0\rangle
\end{equation}
where $S_{Z^{\prime}}=(M_{ \psi} ^2 - M_{Z^{\prime  }}^2 + i \Gamma_{Z^\prime} M_{Z^\prime})^{-1}$ is the propagator of $Z^{\prime}$, and $M_{Z^{\prime}}~(\Gamma_{Z^\prime})$ corresponds to its mass~(decay width).
Here, $f_{\psi}$ and $M_{\psi}$ represent the decay constant and mass of $J/\psi$, respectively, while $\alpha_{em}$ corresponds to the QED fine structure constant. 
Incorporating the interference between amplitudes outlined in Eq.~\eqref{eq4}, we arrive at a first-order approximation of the polarization asymmetry $\alpha_{\text{NP}}$ as:
\begin{equation}
\alpha_{\text{NP}}=\frac{3g_Ag_V}{2\pi\alpha_{em}}\frac{1-r}{(r-1)^2 + y^2 }{\cal F}_0 \propto  \frac{2 |A_{Z^\prime}| }{|A_{\gamma} | } \,,
\end{equation}
where the ratios of $M_{Z^\prime}^2/M_\psi^2$ and $\Gamma_{Z^\prime}/M_{Z^\prime}$ are written as $r$ and $y$, respectively.
It is crucial to emphasize that ${\cal F}_0$ depends only on the ratios of the timelike baryonic form factors, which reduces certain uncertainties.
We adopt ${\cal F}_0=0.67$ from the $^3P_0$ model, which aligns well with experimental measurements, as detailed in Ref.~\cite{Geng:2023yqo}.
Due to the computation of $\Gamma_{Z^\prime}$ requiring a comprehensive knowledge of the effective Lagrangian, which introduces additional unknown coefficients, and the observation that the dependence of $\alpha_{\text{NP}}$ on $\Gamma_{Z^\prime}$ can be safely neglected under the narrow width assumption, we have opted to set $y=0.01$ in our subsequent evaluation.

We are now prepared to evaluate the discovery potential of the $Z^{\prime}$ boson, both within the existing BESIII experiment and in anticipation of future experiments at the STCF.
The total number of events is provided as $N_{\text{event}}=N_{J/\psi}\times{\cal B}_{\Lambda\overline{\Sigma}^{0}+c.c}\times\epsilon$, where $N_{J/\psi}$ represents the number of produced $J/\psi$ particles, ${\cal B}_{\Lambda\overline{\Sigma}^{0}+c.c}$ is given as $2.83\times10^{-5}$ and $\epsilon$ denotes the detector efficiency concerning the considered final states.
For BESIII and STCF experiments, $N_{J/\psi}$ is estimated to be $10^{10}$ and $3.4\times10^{12}$, respectively~\cite{Mangoni:2020nsc,Zeng:2023wqw}.
The detector efficiencies at the BESIII regarding to $\Lambda\overline{\Sigma}^{0}$ and $\overline{\Lambda}\Sigma^{0}$ are $17.6\%$ and $21.7\%$~\cite{BESIII:2012xdg}, respectively.
We take $\epsilon=0.2$ in the following for the sake of simplicity.
We have also adopted a theoretical value of $\alpha_{\text{NP}}=0.02$, which is well within the reach by the BESIII experiment and can be easily surpassed by the STCF.
Based on the anticipated events $N_{\text{event}}$, along with the specified $\alpha_{\text{NP}}$, we conducted simulations of the angular distribution using the Monte Carlo method.

\begin{figure}[htbp]
\centering
\begin{subfigure}[b]{0.45\textwidth}
\centering
\includegraphics[width=\textwidth]{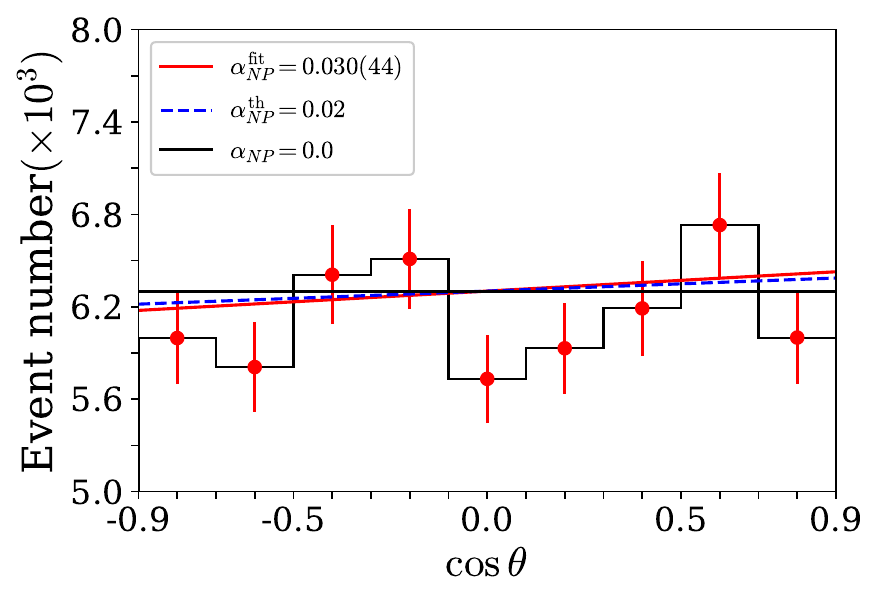}
\caption{BESIII}
\label{sima}
\end{subfigure}
\hfill
\begin{subfigure}[b]{0.47\textwidth}
\centering
\includegraphics[width=\textwidth]{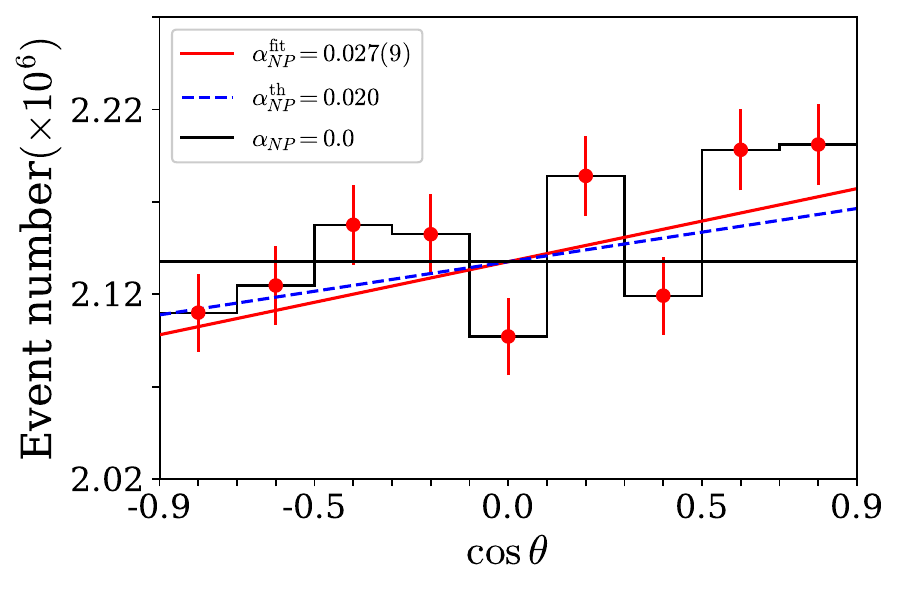}
\caption{STCF}
\label{simb}
\end{subfigure}
\caption{Simulation results for (a) BESIII and (b) STCF, where the systematic uncertainties are taken to be $5\%$ and $1\%$, respectively. }
\label{sim}
\end{figure}

The simulation data are listed in Tab.~\ref{simdata}, where we have taken the endpoints of detector to be $|\cos\theta|=0.9$.
We plot both the simulation and fitting results in Fig.~\ref{sim}.

\begin{table}[htbp]
\caption{The Monte-Carlo simulation data for both BESIII and STCF.}
\label{simdata}
\begin{tabular}{p{2cm}c|c}
\hline
\hline
& \multicolumn{2}{c}{Event numbers} \\
\hline
$\cos\theta$ & BESIII & STCF \\
\hline
(-0.9,-0.7) & $5998\pm77_{stat.}\pm300_{syst.}$ & $2110070\pm1453_{stat.}\pm21101_{syst.}$ \\
(-0.7,-0.5) & $5809\pm76_{stat.}\pm290_{syst.}$ & $2124759\pm1458_{stat.}\pm21248_{syst.}$ \\
(-0.5,-0.3) & $6409\pm80_{stat.}\pm320_{syst.}$ & $2157609\pm1469_{stat.}\pm21576_{syst.}$ \\
(-0.3,-0.1) & $6511\pm81_{stat.}\pm326_{syst.}$ & $2152493\pm1467_{stat.}\pm21525_{syst.}$ \\
(-0.1, 0.1) & $5732\pm76_{stat.}\pm287_{syst.}$ & $2097168\pm1448_{stat.}\pm20972_{syst.}$ \\
( 0.1, 0.3) & $5931\pm77_{stat.}\pm297_{syst.}$ & $2184067\pm1478_{stat.}\pm21841_{syst.}$ \\
( 0.3, 0.5) & $6190\pm79_{stat.}\pm310_{syst.}$ & $2119242\pm1456_{stat.}\pm21192_{syst.}$ \\
( 0.3, 0.7) & $6731\pm82_{stat.}\pm337_{syst.}$ & $2198185\pm1483_{stat.}\pm21982_{syst.}$ \\
( 0.7, 0.9) & $6001\pm77_{stat.}\pm300_{syst.}$ & $2201041\pm1484_{stat.}\pm22010_{syst.}$ \\
\hline
\hline
\end{tabular}
\end{table}

We adopt the minimum $\chi^{2}$ fitting method for the simulation data, where the fit function is given as
\begin{equation}
\chi^{2}=\sum_{\theta}\left(\frac{N_{\text{event}}^{\text{sim}}-N_{\text{event}}^{\text{theory}}(\alpha_{\text{NP}},N_{\text{fit}})}{\text{tot. err}\times N_{\text{event}}^{\text{theory}}}\right)^{2},~\text{tot. err} = \sqrt{\text{sys. err}^2+1/N_{\text{event}} ^{\text{sim}} }
\end{equation}
where the systematic uncertainties in event numbers are assumed to be 5\% at BESIII and 1\% at STCF, and we take ndf=7 in both cases.
The fitted values for $(\alpha_{\text{NP}},N_{\text{fit}})$ are obtained by
\begin{equation}
\frac{\partial\chi^{2}}{\partial\alpha_{\text{NP}}}=0,~\frac{\partial\chi^{2}}{\partial N_{\text{fit}}}=0
\end{equation}
and we obtain $\chi^2/\text{ndf}=1.18,2.27$ for BESIII and STCF, respectively.
The uncertainties are given by the inverse of covariance matrix
\begin{equation}
(V^{-1})_{i j}=\frac{1}{2} \frac{\partial^2 \chi^{2}}{\partial a_i \partial a_j},~V_{ii}=\sigma^{2}_{i}
\end{equation}
where $a_{1,2}$ represent $\alpha_{\text{NP}},N_{\text{fit}}$ respectively, $\sigma_{1}$ is the standard deviation for $\alpha_{\text{NP}}$.

As shown in Fig.~\ref{sim}, the fitted results for BESIII and STCF are $\alpha_{\text{NP}}=0.030\pm0.044$ and $\alpha_{\text{NP}}=0.027\pm0.009$, where the error is mainly statistical.
Take the goodness of fit $\chi^2/\text{ndf}$ into consideration, we obtain the significances are 0.8$\sigma$ and 2.2$\sigma$ for BESIII and STCF respectively.
Furthermore, with the anticipated significant improvements in both statistical and systematic precision at the forthcoming STCF, the prospects for detecting the $Z^\prime$ boson become much more promising, even for smaller values of $\alpha_{\text{NP}}$.

\begin{figure}[htbp]
\centering
\begin{subfigure}[b]{0.45\textwidth}
\centering
\includegraphics[width=\textwidth]{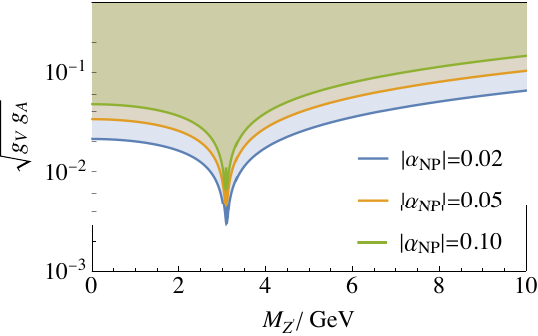}
\caption{General case}
\label{cona}
\end{subfigure}
\hfill
\begin{subfigure}[b]{0.44\textwidth}
\centering
\includegraphics[width=\textwidth]{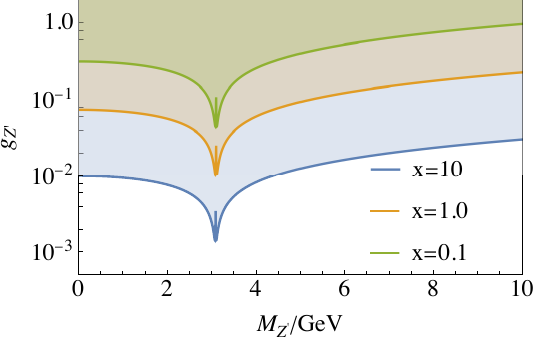}
\caption{$d-xu$ model with $\alpha_{\text{NP}}=0.02$}
\label{conb}
\end{subfigure}
\caption{Coupling-Mass curves for (a) the general case and (b) a specific $Z^{\prime}$ model.}
\label{con}
\end{figure}

Importantly, it should be recognized that such an exploration bears significance even when no significant signal has been found.
In such case, stringent constraints on the gauge coupling of $Z^{\prime}$ relative to its mass are established, taking into account the promising precision of $\alpha_{\text{NP}}$ measurements at the BESIII and STCF.
These constraints are depicted in Fig.~\ref{con}.

In Fig.~\ref{cona}, we consider the model-independent scenario, where the exclusion regions are clearly depicted above the solid lines.
Our constraints on $\sqrt{g_Vg_A}$ span the range of $10^{-2}\sim10^{-1}$, which surpasses the existing bounds established by the CMS experiment~\cite{CMS:2018ucw}. It is worth noting that the mass of the $Z^{\prime}$ boson exerts only a minimal influence on the exclusion curves, with the exception being the vicinity of $M_{J/\psi}$.
For specific models, we can derive constraints on the gauge coupling $g_{Z^{\prime}}$ provided that we have knowledge of the quantum numbers of the $U(1)^{\prime}$ gauge group.
As an illustration, we consider the $d-xu$ model, in which
the couplings between $Z^{\prime}$ and quarks are given as~\cite{ParticleDataGroup:2022pth}
\begin{equation}
g_u^R=-\frac{x}{3}g_{Z^{\prime}},~g_d^R=\frac{1}{3}g_{Z^{\prime}},~g^{L}=0,
\end{equation}
where $x$ can be any rational value.
In Fig.~\ref{conb}, we present the excluded parameter space for various values of $x$, assuming an upper limit of $\alpha_{\text{NP}}$ at 0.02.
Our approach is a valuable complement to other research efforts when studying $Z^{\prime}$ bosons with masses below 10 GeV.

%{\color{red}
%We have also considered relaxing the leptophobic constraint of our model by adding $g_eZ^\prime_\mu\bar{e}\gamma^{\mu}e$ to the effective Lagrangian.
%As a result, we could now extend our consideration to the parity violation in $e^{+}e^{-}\to\Lambda\overline{\Sigma}^{0}$, which is caused by the interference between $e^{+}e^{-}\to Z^{\prime}\to\Lambda\overline{\Sigma}^{0}$ and $e^{+}e^{-}\to J/\psi\to\Lambda\overline{\Sigma}^{0}$.
%Unfortunately, such parity violation is severely suppressed by $g_{e}^{2}/(4\pi\alpha_{em})$, where the upper bound of the coupling $g_{e}$ in this mass range is around $10^{-4}$.
%Therefore, such parity violation is far beyond the current sensitivity range of BESIII experiments.
%}

% \section{Conclusion}

In conclusion, we have explored the new possibility of discovering the $Z^{\prime}$ boson with a mass below 10 GeV, a range currently accessible at the BESIII. 
Our simulations indicate that these signals could be detected at the BESIII, provided that the systematic uncertainty is further reduced. 
There is also a potential for improved signal detection at the future STCF.
If no clear signal emerges, we can still derive useful information by establishing general constraints on the couplings of the $Z^{\prime}$ boson to quarks, typically falling within the range of $10^{-2}$ to $10^{-1}$, regardless of the $Z^{\prime}$ mass.
Our approach offers a competitive and complementary method for hunting down the $Z^{\prime}$ boson with a mass below 10 GeV.
Even in less favorable scenarios, it can still make valuable contributions to the constraints on  the couplings of the $Z^{\prime}$ boson to quarks in the low mass range.

\section*{Acknowledgments}
We would like to express our sincere appreciation to Prof. Haibo Li and Baician Ke for providing valuable information during the development of this work.
This work is supported in part by the National Key Research and Development Program of China under Grant No. 2020YFC2201501 and  the National Natural Science Foundation of China (NSFC) under Grant No. 12347103 and 12205063.

\end{document}